\documentclass[aps,showpacs,onecolumn,floatfix,amsmath,amssymb,nofootinbib]{revtex4-1}

\usepackage{amssymb}
\usepackage{amsmath}
\usepackage{epsfig}
\usepackage{graphicx}
\usepackage{color}
\usepackage{latexsym}
\usepackage{bm,latexsym}
\usepackage[mathscr]{eucal}
\usepackage{float}
\usepackage{pbox}

\begin{document}

\title{ Ellis wormhole  without a phantom scalar field }

\author{Pedro Ca\~nate$^{1}$}
\email{pcanate@fis.cinvestav.mx, pcannate@gmail.com}

\author{Joseph Sultana$^{2}$}
\email{joseph.sultana@um.edu.mt}

\author{Demosthenes Kazanas$^{3}$}
\email{demos.kazanas-1@nasa.gov}

\affiliation{$^{1}$Departamento de F\'isica, Centro de Investigaci\'on y de Estudios Avanzados
del I.P.N.,\\ Apdo. 14-740, 07000 Mexico City, Mexico.\\
$^{2}$Department of Mathematics, Faculty of Science,
University of Malta, Msida, Malta\\
$^{3}$Astrophysics Science Division, NASA/Goddard Space Flight Center, Greenbelt, Maryland 20771, USA}

\begin{abstract}
In this paper, we present an exact solution for $(3+1)$-dimensional Einstein-scalar-Gauss-Bonnet theory (EsGB) in electrovacuum. The solution is characterized by only one parameter, $Q$, which in general can be associated with the electromagnetic field and the scalar field.
 We show that the solution corresponds to a charged wormhole with a throat at the region $r = |Q|$ and is also supported by a real scalar field having a positive kinetic term.  We show that the solution belongs to the most general class of solutions known as Ellis wormholes but without the need for ``exotic matter" or a phantom scalar field.
\end{abstract}



\maketitle

\section{Introduction}
An interesting phenomenon which arises from the geometrical description of gravity is the existence of wormholes \cite{visser95}. These involve a topological spacetime configuration in the form of a short-cut between distant points or regions in spacetime. The idea originates from the work of Einstein and Rosen in 1935, with their solution known as the Einstein-Rosen bridge \cite{einstein35}, which is basically the maximally extended Schwarzschild solution. However it quickly turned out \cite{kruskal60} that the ``throat" of such a wormhole is dynamic and hence nontraversable, meaning that its radius expands to a maximum and quickly contracts to zero so fast that even a photon cannot pass through. Following this, interest in wormholes was revived by the work of Morris and Thorne in 1988 \cite{morris88}, who obtained traversable wormholes without event horizons, and showed that the throat of these wormholes can be kept open by some form of ``exotic" matter \cite{morris88-2} having negative energy density, and whose energy momentum tensor violates the null-energy condition (NEC). An example of this exotic type of matter may be represented by a phantom scalar field having a negative kinetic term in the gravitational action. This means that at a classical level traversable Morris-Thorne wormholes are forbidden in general relativity because all know types of matter satisfies the null-energy condition. Examples of traversable Morris-Thorne wormholes are exhibited in \cite{wormholes}. However recently, using ideas from guage/gravity duality, it has been shown that quantum matter fields can provide the necessary negative energy to keep the throat of the wormhole open and thus achieve traversable wormholes. This was first done for asymptotically anti-de Sitter wormholes \cite{gao17,maldacena17,maldacena18} and then generalized to the asymptotically flat case \cite{maldacena18-2,fu19}. In all cases it was found that given two points, one on each side of the wormhole's throat, it will take longer to go from one point to the other through the wormhole, instead of around it. More recently \cite{horowitz19} it was shown that wormholes can also be produced through a quantum tunneling event, where a backreaction from quantum fields can make these wormholes traversable.\\[2mm]
The need for such exotic matter in traversable wormholes can be circumvented in modified theories of gravity which contain higher curvature corrections in their gravitational actions. In these theories the effect of these higher order terms leads to the violation of the null-energy condition and can thus mimic the exotic matter contribution required for traversable wormholes. So for example, traversable wormholes without matter sources have been obtained in quadratic gravity \cite{duplessis15} and in Chern-Simons modified gravity \cite{nascimento17}. Other wormhole solutions without exotic matter or phantom fields have been obtained in Einstein-Cartan gravity \cite{bronnikov16}, conformal Weyl gravity \cite{lobo08}, conformal scalar-tensor theories \cite{barcelo99}, Lovelock gravity \cite{zangeneh15}, Horndeski gravity \cite{korolev14}, bumblebee gravity \cite{ovgun19}, $f(R,\phi)$ gravity \cite{zubair18} and $f(T)$ gravity \cite{bohemer12}. It should be pointed out that many of these wormhole solutions were found to be unstable. For example a no-go theorem for wormholes has been proved for Horndeski gravity \cite{evseev18} which basically states that there are no stable static and spherically symmetric wormhole solutions in this theory (although this may not remain the case for the more general beyond Horndeski theories as shown in Ref. \cite{mironov18}).\\[2mm]
One of the first, and simplest examples of traversable wormholes in general relativity is the static and spherically symmetric Ellis wormhole \cite{ellis73} whose energy momentum tensor can be represented by a massless phantom scalar field. This solution has been extensively studied, and its properties like gravitational lensing \cite{EllisLensing}, quasinormal modes \cite{QNMEllis}, shadows \cite{EllisShadows} and stability \cite{EllisStability} have all been thoroughly investigated. Recently \cite{bronnikov13} it was shown that the source of the Ellis wormhole can also be represented by a perfect fluid with negative energy density and a source-free radial electric or magnetic field. In this paper we show that the Ellis wormhole is also an exact solution of the Einstein-scalar-Gauss-Bonnet (EsGB) theory in an electrovacuum, with the same source-free radial electric field used in Ref. \cite{bronnikov13}, but now with a real scalar field having a positive kinetic term. Einstein-scalar-Gauss-Bonnet theories represent an interesting class of alternative theories of gravity, in which the Einstein-Hilbert action is modified by including a scalar field that is nonminimally coupled to the Gauss-Bonnet term. Although the Gauss-Bonnet term includes quadratic curvature components, the resulting field equations arising from the EsGB action are of second order, and therefore avoid the Ostrogradski instability and ghosts. For  this  theory novel black holes have been  numerically  derived in \cite{Antoniou18}. In our case the coupling between the scalar field and the Gauss-Bonnet term is what creates the necessary violation of the null-energy condition required for the traversable Ellis wormhole, and so in a sense the effective negative energy density comes from the geometry itself instead of the matter source as in GR. Recently \cite{antoniou19} a number of novel wormhole solutions in EsGB theory have been obtained numerically for several coupling functions. Other traversable wormhole solutions have been obtained earlier \cite{kanti11} in Einstein-dilaton-Gauss-Bonnet theory \cite{kanti96}, which involve an exponential coupling between the scalar field representing the dilaton and the Gauss-Bonnet term.\\[2mm]
In the next section we obtain the field equations for the EsGB gravity coupled to Maxwellian electrodynamics. Then in Sec. III we show that the Ellis wormhole is a solution of these field equations for a radial source-free electric field and a real scalar field. We also consider the limiting case of a vanishing scalar field. This is followed by the conclusion and discussion. In this paper we use units where $G = c = 1$.

\section{Einstein-scalar-Gauss-Bonnet gravity coupled to Maxwell electrodynamics}
In this section we shall briefly describe the dynamical equations of EsGB theory coupled to Maxwell electrodynamics. This theory is  described by the following action functional:
\begin{equation}\label{actionL}
S[g_{ab},\phi,A_{a}] = \int d^{4}x \sqrt{-g} \left\{ \frac{1}{16\pi}\left(R - \frac{1}{2}\partial_{\mu}\phi\partial^{\mu}\phi + \boldsymbol{f}(\phi) R_{_{GB}}^{2} \right) + \frac{\tilde{\kappa}}{4\pi} F   \right\},
\end{equation}
where the EsGB contribution is due to the Ricci scalar $R$, the quadratic Gauss-Bonnet term  $R_{_{GB}}^{2}$ defined
by $R_{_{GB}}^{2} = R_{\alpha\beta\mu\nu}R^{\alpha\beta\mu\nu} - 4R_{\alpha\beta}R^{\alpha\beta} + R^{2}$, and the scalar field $\phi$ which is coupled to the GB term through a general coupling function $\boldsymbol{f}(\phi)$. Whereas, the electromagnetic contribution is due to the Lagrangian of the electromagnetic theory, $\tilde{\kappa} F$, where $\tilde{\kappa}$ is a parameter of the theory, and $F$ is the electromagnetic field invariant $F=\frac{1}{4}F_{ab}F^{ab}$ with $F_{ab}$ the electromagnetic field tensor defined in terms of the electromagnetic potential $A_{a}$ by $F_{ab} = \partial_{a} A_{b} - \partial_{b} A_{a}$. \\[2mm]
The Einstein-scalar-Gauss-Bonnet-Maxwell equations arising from varying this action with respect to the gravitational field,
\begin{equation}\label{EinsteinEqs}
G_{a}{}^{b} = 8\pi (E_{a}{}^{b})\!_{_{_{S\!G\!B}}} + 8\pi (E_{a}{}^{b})\!_{_{_{ E \! D}}},
\end{equation}
where the scalar-field-Gauss-Bonnet (SGB) effective energy-momentum tensor $(E_{a}{}^{b})\!_{_{_{S\!G\!B}}}$, and the electromagnetic stress-momentum tensor $(E_{a}{}^{b})\!_{_{_{ E \! D}}}$, are given respectively by
\begin{eqnarray}
&&8\pi (E_{\alpha}{}^{\beta})\!_{_{_{S\!G\!B}}} = -\frac{1}{4}(\partial_{\mu}\phi \partial^{\mu}\phi)\delta_{\alpha}{}^{\beta} + \frac{1}{2}\partial_{\alpha}\phi \partial^{\beta}\phi - \frac{1}{2}( g_{\alpha\rho} \delta_{\lambda}{}^{\beta} + g_{\alpha\lambda} \delta_{\rho}{}^{\beta})\eta^{\mu\lambda\nu\sigma}\tilde{R}^{\rho\xi}{}_{\nu\sigma}\nabla_{\xi}\partial_{\mu}\boldsymbol{f}(\phi), \\ \label{E_GB}
&&8\pi (E_{\alpha}{}^{\beta})\!_{_{_{ E \! D}}} =  2 \tilde{\kappa} \left( \frac{1}{4}F_{\mu\nu}F^{\mu\nu} \delta_{\alpha}{}^{\beta} -  F_{\alpha\mu}F^{\beta\mu}\right), \label{E_NLED}
\end{eqnarray}
with $\tilde{R}^{\rho\gamma}{}_{\mu\nu} = \eta^{\rho\gamma\sigma\tau}R_{\sigma\tau\mu\nu} = \epsilon^{\rho\gamma\sigma\tau}R_{\sigma\tau\mu\nu}/\sqrt{-g}$. On the other hand, the variation with respect to the electromagnetic potential $A_{a}$ yields the electromagnetic field equations

\begin{equation}\label{EM_Eqs}
\nabla_{\alpha}( F^{\alpha\beta} ) = 0,
\end{equation}
while the variation with respect to the scalar field yields the equation of motion for $\phi$,

\begin{equation}\label{scalar_Eq}
\nabla^{2}\phi + \dot{\boldsymbol{f}}(\phi) R_{_{GB}}^{2} = 0,
\end{equation}
where the dot $\dot{\boldsymbol{f}}$ denotes the derivative  of $\boldsymbol{f}$ with respect to the scalar field.\\[2mm]
Our aim is to find a solution of the set of Eqs. (\ref{EinsteinEqs}), (\ref{EM_Eqs})  and (\ref{scalar_Eq})
that describe a static and spherically symmetric, asymptotically flat charged wormhole solution with a nontrivial scalar field. Therefore, we will assume that the scalar field is static and spherically symmetric, $\phi = \phi(r)$, and also that the metric takes the static and spherically symmetric form,

\begin{equation}\label{SSSmet}
ds^{2} =  - e^{ A(r) }dt^{2} + e^{ B(r) }dr^{2}  + r^{2}(d\theta^{2}  + \sin^{2}\theta d\varphi^{2}),
\end{equation}
with $A = A(r)$ and $B = B(r)$ being unknown functions depending only on $r$. \\[2mm]
As regards to the electromagnetic field, given that the spacetime is static and spherically symmetric, we can assume that the only non null electromagnetic field tensor component is $F_{tr}=F_{tr}(r)$; i.e., we shall restrict ourselves to an electromagnetic field given by

\begin{equation}\label{f_ab_E}
F_{\alpha\beta} = \mathcal{E}(r)\left( \delta^{t}_{\alpha}\delta^{r}_{\beta} - \delta^{r}_{\alpha}\delta^{t}_{\beta} \right).
\end{equation}
In this way, the general solution of the Eq. (\ref{EM_Eqs}) is given by

\begin{equation}\label{fabSOL}
\nabla_{\alpha}(  F^{\alpha\beta} ) = 0 \quad \Rightarrow \quad  \partial_{r}( \sqrt{-g}  F^{rt}) = 0 \quad \Rightarrow \quad F^{\alpha\beta} = \frac{ \mathcal{Q} }{ \sqrt{- r^{4} g_{tt}g_{rr} }  }\delta_{r}{}^{\alpha}\delta_{t}{}^{\beta} \quad\Rightarrow \quad F = - \frac{ \mathcal{Q}^{2} }{ 2r^{4} },
\end{equation}
where $\mathcal{Q}$ is an integration constant.

$$\textup{ {\bf Non-null Einstein tensor components} }$$\\[2mm]
For the line element (\ref{SSSmet}), the only non-null Einstein tensor components are given by

\begin{eqnarray}
&& G_{t}{}^{t} = \frac{ e^{ -B} }{ r^{2} }\left( -rB' - e^{ B} + 1 \right), \quad\quad  G_{r}{}^{r} = \frac{ e^{ -B} }{ r^{2} }\left( rA' - e^{ B} + 1 \right), \label{GttyGrr}\\
&& G_{\phi}{}^{\phi} = G_{\theta}{}^{\theta} = \frac{ e^{ -B} }{ 4r } \left( rA'^{2} - rA'B' + 2rA'' + 2A' - 2B' \right),
\label{GththyGphph}
\end{eqnarray}
where the prime denotes the derivative with respect to the radial coordinate $r$.

$$\textup{ {\bf Non-null SGB effective-energy-momentum tensor components} }$$\\[2mm]
For the line element (\ref{SSSmet}), the only nonvanishing SGB effective-energy-momentum  tensor components are given by

\begin{eqnarray}
&& 8\pi (E_{t}{}^{t})\!_{_{_{S\!G\!B}}} = -\frac{ e^{ -2B} }{ 4r^{2} }\left\{  \left[r^{2}e^{B} + 16(e^{B} - 1)\ddot{\boldsymbol{f}} \right]\phi'^{2} - 8[ (e^{B} - 3)B'\phi' - 2(e^{B} - 1)\phi'']\dot{\boldsymbol{f}} \right\}, \label{GBtt} \\
&& 8\pi (E_{r}{}^{r})\!_{_{_{S\!G\! B}}} = \frac{ e^{ -B} \phi'}{ 4 } \left[ \phi'  - \frac{8(e^{B} - 3)e^{-B}A'\dot{\boldsymbol{f}} }{r^{2}}  \right],  \label{GBrr} \\
&& 8\pi (E_{\theta}{}^{\theta})\!_{_{_{S\!G\! B}}} = 8\pi (E_{\phi}{}^{\phi})\!_{_{_{S\!G\! B}}} = -\frac{ e^{ -2B} }{ 4r } \left\{ ( re^{B} - 8A'\ddot{\boldsymbol{f}})\phi'^{2} - 4\left[ (A'^{2} + 2A'')\phi' + (2\phi'' - 3B'\phi')A'\right]\dot{\boldsymbol{f}} \right\}.
\end{eqnarray}

$$\textup{ {\bf Non-null electromagnetic stress-energy tensor components} }$$\\[2mm]
For the line element (\ref{SSSmet}), and electromagnetic field tensor (\ref{f_ab_E}), the only non-null electromagnetic stress-energy tensor components are given by

\begin{eqnarray}
&& 4\pi (E_{t}{}^{t})\!_{_{_{ E \! D}}} = 4\pi (E_{r}{}^{r})\!_{_{_{ E \! D}}} = - \tilde{\kappa} F,
\label{ttyrr}\\
&& 4\pi (E_{\theta}{}^{\theta})\!_{_{_{ E \! D}}} = 4\pi (E_{\phi}{}^{\phi})\!_{_{_{ E \! D}}} = \tilde{\kappa} F. \label{ththyphph}
\end{eqnarray}

\subsection{ Field equations }
The EsGB-Maxwell equations for the static and spherically symmetric space-time take the form,

\begin{eqnarray}
&& G_{t}{}^{t} = 8\pi E_{t}{}^{t} \quad \Rightarrow \quad   4e^{B}\left( rB' + e^{ B} - 1 \right) =  \left[ r^{2}e^{B} +16(e^{ B}-1)\ddot{\boldsymbol{f}} \right]\phi'^{2} - 8  \left[ (e^{ B} - 3)B'\phi' - 2(e^{ B} - 1)\phi'' \right]\dot{\boldsymbol{f}} \nonumber \\
&& \hskip6.6cm + \!\! \quad  8 \tilde{\kappa} r^{2}e^{2B} F,
\label{Eqt}\\
&&G_{r}{}^{r} = 8\pi E_{r}{}^{r} \quad \Rightarrow \quad   4e^{B}\left( -rA' + e^{ B} - 1 \right) =   -r^{2}e^{B}\phi'^{2} + 8(e^{ B} - 3)A'\phi'\dot{\boldsymbol{f}} + 8 \tilde{\kappa} r^{2}e^{2B} F,
\label{Eqr}\\
&&G_{\theta}{}^{\theta} = 8\pi E_{\theta}{}^{\theta} \quad \Rightarrow \quad  e^{B}\left\{ rA'^{2} - 2B' + (2 - rB')A'   + 2rA''   \right\} = -re^{B}\phi'^{2} + 8A'\ddot{\boldsymbol{f}}\phi'^{2} \nonumber \\
&& \hskip6.6cm + \!\! \quad 4  \left[  (A'^{2} + 2A'')\phi' + (2\phi'' - 3B'\phi')A' \right]\dot{\boldsymbol{f}} +  8 \tilde{\kappa} re^{2B}F,
\label{Eqte}
\end{eqnarray}
whereas the equation of motion for the scalar field becomes

\begin{equation}\label{phi2}
2r\phi'' + (4 + rA' - rB')\phi' + \frac{4e^{-B}\dot{\boldsymbol{f}}}{r} \left[ (e^{B} - 3)A'B' - (e^{B} - 1)(2A'' + A'^{2})\right] = 0.
\end{equation}
The solution of the Maxwell equations is given by (\ref{fabSOL}). Therefore, the electromagnetic invariant will be $F = - \frac{ \mathcal{Q}^{2} }{ 2r^{4} }.$

In this case we see that Reissner-Nordstr\"om metric, $e^{A(r)} = e^{-B(r)} = 1 - \frac{2 M }{r} + \frac{ Q^{2} }{r^{2}}$, is a solution of the system of equations, (\ref{fabSOL}), (\ref{Eqt}), (\ref{Eqr}), (\ref{Eqte}) and (\ref{phi2}), with, $\phi(r)$ = constant, $\boldsymbol{f}(\phi)$ = constant,  $\tilde{\kappa} = - (4\pi\epsilon)^{2}$ and $F = - \frac{ Q^{2}  }{ 32\pi^{2}\epsilon^{2} r^{4} }$. We can see that if the electromagnetic field is turned off, $F(r)=0$, then the system of equations (\ref{fabSOL}), (\ref{Eqt}), (\ref{Eqr}), (\ref{Eqte}) and (\ref{phi2}), reduce to that for pure EsGB gravity as presented in Ref. \cite{Antoniou18}.

\section{Four-dimensional traversable charged wormhole solutions in Einstein-scalar-Gauss-Bonnet theory}
The canonical metric for a (3+1)-dimensional static and spherically symmetric wormhole (SSS-WH) is given by \cite{morris88,morris88-2}
\begin{equation}\label{WH_Thorne}
ds^{2} =  -  e^{2 \Phi(r)} dt^{2} + \frac{dr^{2}}{ 1  - \frac{b(r)}{r} }  + r^{2} \left( d\theta^{2} + \sin^{2}\theta d\varphi^{2}\right),
\end{equation}
where $\Phi(r)$ and $b(r)$ are functions of the radial coordinate $r$. The function, $\Phi(r)$, is called the redshift function, for it is related to the gravitational redshift, while $b(r)$ is referred to as the shape function.  The radial coordinate has a range that increases from a minimum value at $r_{0}$, corresponding to the wormhole throat, in which $b(r_{0}) = r_{0}$, to $r \rightarrow \infty$. Thus, the radial coordinate $r$ has a special geometric significance, where $4\pi r^2$  is the area of a sphere centered on the wormhole throat.
On the other hand, for the wormhole to be traversable, one must demand the absence of event horizons, which are identified as the surfaces where $e^{2\Phi(r)} \rightarrow 0$.  If $\Phi(r)$ is a continuous, nonvanishing and finite function in the whole range of $r$,  $r\in[r_{0},\infty)$, there will not be any horizons. Moreover, a fundamental property of traversable wormholes is the fulfilment of the flaring out condition \cite{morris88,morris88-2}, which is deduced from the mathematics of embedding, and is given by $(b - rb')/b^{2} > 0$. Note that at the throat,  $b(r_{0}) = r_{0},$  the flaring out condition reduces to $b'(r_{0})<1$. The condition $(1 - b/r) \geq 0$ is also imposed for all values of $r$. \\[2mm]
In general relativity (GR) there is a connection between the flaring out condition and
the null energy condition (NEC). This connection can be determined considering a null vector, $\boldsymbol{n}$, given by
$\boldsymbol{n} = \left( e^{-\Phi(r)}, \sqrt{ 1  - \frac{b(r)}{r} }, 0, 0 \right) = n^{\alpha}\partial_{\alpha}$.  Where the set $\{ n^{\alpha} \}$, with $\{\alpha=t, r, \theta, \phi\}$, are the components of the null vector in the  basis vectors $\{ \partial_{t}, \partial_{r}, \partial_{\theta}, \partial_{\phi} \}$
such that $dx^{\alpha}(\partial_{\beta}) = \partial_{\beta}(dx^{\alpha}) = \delta^{\alpha}_{\beta}$. We can see that  $\boldsymbol{n}$ is a null vector since $ds^{2}( \boldsymbol{n}, \boldsymbol{n}) = g_{_{\alpha\beta}}n^{\alpha}n^{\beta} = 0$.  Then, contracting the Einstein tensor with $\boldsymbol{n}$ we find

\begin{equation}
G_{\alpha\beta} n^{\alpha} n^{\beta} = e^{-2\Phi(r)}G_{tt} + \left( 1  - \frac{b(r)}{r} \right)G_{rr} = e^{-2\Phi(r)}g_{tt}G_{t}{}^{t} + \left( 1  - \frac{b(r)}{r} \right)g_{rr}G_{r}{}^{r} = - G_{t}{}^{t} + G_{r}{}^{r}
\end{equation}
Therefore, by identifying; $e^{A(r)} = e^{2\Phi(r)}$ and $e^{B(r)} = \frac{1}{1 - \frac{b(r)}{r} }$, and using (\ref{GttyGrr}), yields

\begin{equation}
G_{\alpha\beta} n^{\alpha} n^{\beta} = \frac{2}{r} \left( 1 - \frac{b(r)}{r}  \right) \Phi'(r) - \frac{1}{r^{3}} \left[  b(r) - rb'(r) \right].
\end{equation}
Evaluating the rhs of the above equation at the wormhole throat, $r=r_{0}$, and using the fact that $\Phi(r)$ is continuous and finite everywhere, the quantity $G_{(\alpha)(\beta)} n^{(\alpha)} n^{(\beta)}|_{r=r_{0}}$ reduces to

\begin{eqnarray}\label{flout}
 G_{\alpha\beta}n^{\alpha}n^{\beta}  \Big|_{r=r_{0}} =  -\frac{1}{r^{3}_{0}} \left[ b( r_{0}) - r_{0}b'(r_{0})  \right] =\frac{1}{r^{2}_{0}} \left[ b'
(r_{0})- 1  \right]< 0, \quad  {\rm  since } \quad  b'(r_{0}) < 1.
\end{eqnarray}
Then using Einstein field equations $G_{\alpha\beta} = \kappa T_{\alpha\beta}$  at the wormhole's throat $r_{0}$, we get

\begin{equation}
T_{\alpha\beta}n^{\alpha}n^{\beta}\!\Big|_{r=r_{0}} < 0.
\end{equation}
This implies that in GR the NEC (which establishes that for any null vector $n^{\alpha}$, $T_{\alpha\beta}n^{\alpha}n^{\beta}\geq0$), is violated for a SSS traversable wormhole spacetime.
Therefore, in GR the fulfillment of the flaring out condition, which is considered fundamental for a traversable wormhole, implies the existence of exotic matter (i.e., matter whose energy momentum tensor violates the NEC). 
Now, with regards to EsGB-Maxwell, the field equations (\ref{EinsteinEqs}) have an analogous structure to those in GR with a effective energy-momentum tensor given by $E^{(e\!f\!f)}_{\alpha\beta} =  (E_{\alpha\beta})\!_{_{_{S\!G\!B}}} +  (E_{\alpha\beta})\!_{_{_{E \! D}}}$. This motivates us to calculate $E^{(e\!f\!f)}_{\alpha\beta}n^{\alpha}n^{\beta}$ using the null vector $\boldsymbol{n} = \left( e^{-\Phi(r)}, \sqrt{ 1  - \frac{b(r)}{r} }, 0, 0 \right)$. Thus, one finds that
\begin{equation}
E^{(e\!f\!f)}_{\alpha\beta}n^{\alpha}n^{\beta} = (E_{\alpha\beta})\!_{_{_{S\!G\!B}}}n^{\alpha}n^{\beta} +  (E_{\alpha\beta})\!_{_{_{ E \! D}}}n^{\alpha}n^{\beta} =    \left[ (E_{r}{}^{r})\!_{_{_{S\!G\!B}}} - (E_{t}{}^{t})\!_{_{_{S\!G\!B}}}  \right] + \left[ (E_{r}{}^{r})\!_{_{_{E \! D}}} - (E_{t}{}^{t})\!_{_{_{ E \! D}}}    \right]
\end{equation}
From (\ref{ttyrr}) it follows that $(E_{r}{}^{r})\!_{_{_{ E \! D}}} - (E_{t}{}^{t})\!_{_{_{ E \! D}}}  = 0$; and hence,

\begin{equation}\label{E_GB_null}
E^{(e\!f\!f)}_{\alpha\beta}n^{\alpha}n^{\beta} = (E_{r}{}^{r})\!_{_{_{S\!G\!B}}} - (E_{t}{}^{t})\!_{_{_{S\!G\!B}}}
\end{equation}
Taking advantage of the structure of the EsGB field equations (\ref{EinsteinEqs}) and using (\ref{flout}), we conclude that in EsGB-Maxwell theory, for a traversable wormholes spacetime (\ref{WH_Thorne}), the flaring out condition (\ref{flout}) implies that

\begin{equation}\label{E_GB_null_vio}
E^{(e\!f\!f)}_{\alpha\beta}n^{\alpha}n^{\beta}\!\Big|_{r=r_{0}} = \left[ (E_{r}{}^{r})\!_{_{_{S\!G\!B}}} - (E_{t}{}^{t})\!_{_{_{S\!G\!B}}} \right]\!\Big|_{r=r_{0}} < 0.
\end{equation}
Thus, we see that for a SSS $(3+1)$-dimensional traversable wormhole solution in EsGB-Maxwell, to keep the wormhole's throat open, only the sGB term is necessary. Now we are ready to present an exact solutions in this system. 

\subsection{ Traversable electrovacuum wormhole solution in Einstein-scalar-Gauss-Bonnet theories }\label{whEsGBMax}
The Einstein-scalar-Gauss-Bonnet-Maxwell equations are solved by the following static and spherically symmetric line element:

\begin{equation}\label{NewSolution_maxwell}
ds^{2} =  - dt^{2} + \frac{dr^{2}}{ 1 - \frac{Q^{2}}{r^{2}} }  + r^{2}(d\theta^{2}  + \sin^{2}\theta d\varphi^{2}),
\end{equation}
which is characterized by the electromagnetic field tensor given by (\ref{f_ab_E}), for which the electromagnetic invariant is

\begin{equation}\label{Fsol}
F(r) = - \frac{ (Q^{2} + Q^{2}\!\!\!_{_{s}}) }{ 32\pi^{2}\epsilon^{2} r^{4} },
\end{equation}
while the parameter $\tilde{\kappa}$ in the electromagnetic Lagrangian, $\tilde{\kappa}F$, is given by

\begin{equation}\label{kappa}
\tilde{\kappa} =  - (4\pi\epsilon)^{2}.
\end{equation}
In order to guarantee the consistency of the Eqs. (\ref{fabSOL}) and (\ref{Fsol}), we require that $\mathcal{Q}^{2} = (Q^{2} + Q^{2}\!\!\!_{_{s}})/(4\pi\epsilon)^{2}$.\\[2mm]
On the other hand, for the sGB contribution, this solution is characterized by the scalar field given by

\begin{equation}\label{phi_sol}
\phi(r) =  \frac{ 2 Q\!_{_{s}} }{ Q }\tan^{-1}\left( \frac{ \sqrt{r^{2} - Q^{2}} }{Q} \right),
\end{equation}
On the other hand, the coupling function $\boldsymbol{f}(\phi)$, which characterizes the EsGB theory, is given by the function,

\begin{equation}\label{cfunct}
\boldsymbol{f}(\phi) = \frac{ ( Q^{2} + Q^{2}\!\!\!_{_{s}} ) \left\{ \cos\left( \frac{Q}{2Q\!_{_{s}}}\phi \right)  - \cos^{3}\left( \frac{Q}{2Q\!_{_{s}}}\phi \right) \ln\!\left[ \cos^{4}\left( \frac{Q}{2Q\!_{_{s}}}\phi \right) \right]    -  \left[ 1 + 2\cos^{2}\left( \frac{Q}{2Q\!_{_{s}}}\phi \right) \right]\frac{Q}{Q\!_{_{s}}} \phi\sin\left( \frac{Q}{2Q\!_{_{s}}}\phi\right)     \right\}    }{  12\cos^{3}\left( \frac{Q}{2Q\!_{_{s}}}\phi \right)          }.
\end{equation}
The first, and second, derivative of $\boldsymbol{f}(\phi)$ with respect
to $\phi$, are

\begin{equation}\label{cd_funct}
\dot{\boldsymbol{f}}(\phi) = -\frac{ (Q^{2} + Q^{2}\!\!\!_{_{s}} ) Q^{2} \phi }{ 8  Q^{2}\!\!\!_{_{s}} }  \sec^{4}\left(  \frac{Q}{2Q\!_{_{s}}}\phi \right), \quad
\ddot{\boldsymbol{f}}(\phi) = -\frac{ (Q^{2} + Q^{2}\!\!\!_{_{s}} ) Q^{2}  }{ 8  Q^{3}\!\!\!_{_{s}} } \frac{ \left( 2Q\phi\sin\left(  \frac{Q}{2Q\!_{_{s}}}\phi \right)  + Q\!_{_{s}} \cos\left(  \frac{Q}{2Q\!_{_{s}}}\phi \right) \right)  }{\cos^{5}\left(  \frac{Q}{2Q\!_{_{s}}}\phi \right)}
\end{equation}
The throat of the WH solution (\ref{NewSolution_maxwell}), occurs at $r=r_{0} =|Q|$. We can see that in entire wormhole domain $r \in [|Q|, \infty)$, the scalar field $\phi$ in (\ref{phi_sol}) is real. Its derivative with respect to $r$, yields
\begin{equation}\label{phi_p}
\phi' = \frac{ 2 Q\!_{_{s}} }{ r \sqrt{ r^{2} - Q^{2} }},
\end{equation}
and so the quantity $\phi'$ diverges at the wormhole throat, $r=|Q|$.  However, in contrast to the scalar field $\phi$, the quantity $\phi'$ is not a scalar field invariant. In fact, starting from $\phi' = \nabla_{r} \phi$ we can build an invariant quantity, $\nabla_{\alpha}\phi \nabla^{\alpha}\phi$, which is given by

\begin{equation}
\nabla_{\alpha}\phi \nabla^{\alpha}\phi  = g^{\alpha\beta}\nabla_{\alpha}\phi \nabla_{\beta}\phi= g^{rr}(\nabla_{r}\phi)^{2} = g^{rr}\phi'^{2} = \frac{ 4 Q^{2}\!\!\!_{_{s}} }{ r^{2} \left( r^{2} - Q^{2} \right)}\left(1 - \frac{Q^{2}}{r^{2}} \right) = \frac{ 4 Q^{2}\!\!\!_{_{s}} }{ r^{4} }.
\end{equation}
This is regular everywhere in the wormhole domain.\\[2mm]
A way to establish that the spacetime is regular in the wormhole domain, $r\in[ |Q| , \infty)$, is done by evaluating the curvature invariants $R$, $R_{\alpha\beta}R^{\alpha\beta}$ and $R_{\alpha\beta\mu\nu}R^{\alpha\beta\mu\nu}$ for the metric (\ref{NewSolution_maxwell}) and determine that those invariants behave well in the region of interest, $r\in$ [$|Q|$,$\infty$). This yields

\begin{eqnarray}
&&R = -\frac{ 2Q^{2} }{ r^{4} },\\
&&R_{\alpha\beta}R^{\alpha\beta} = \frac{ 4Q^{4} }{ r^{8} },\\
&&R_{\alpha\beta\mu\nu}R^{\alpha\beta\mu\nu} = \frac{ 12Q^{4} }{ r^{8} }.
\end{eqnarray}
Therefore, we establish that they are all regular everywhere in the wormhole range. Thus at $r=r_{0}=|Q|$ the singularity appearing in (\ref{NewSolution_maxwell}) and (\ref{phi_p}) among others non invariant quantities, like for instance $\phi''= \frac{ 2 Q\!_{_{s}} (Q^{2} - 2r^{2}) }{ r^{2} \left( r^{2} - Q^{2} \right)^{\frac{3}{2} } }$, is only a coordinate singularity describing the existence of the wormhole throat. \\[2mm]
Turning back to our WH solution (\ref{NewSolution_maxwell}), we can see that the redshift and shape functions, are given respectively by $\Phi(r) = \mbox{constant}$ and $b(r) = \frac{Q^{2}}{r}$. Thus, we can see that

\begin{equation}
\frac{(b - rb_{,r})}{b^{2}} = \frac{2r}{Q^{2}} > 0 \quad\quad \textup{ and } \quad\quad  b'(r_0) = -1.
\end{equation}
Thus the solution (\ref{NewSolution_maxwell}) fulfills the flaring out condition and hence represents a traversable wormhole interpretation. As regards to the condition (\ref{E_GB_null_vio}), evaluating (\ref{E_GB_null}) for the line element (\ref{NewSolution_maxwell}), scalar field (\ref{phi_sol}) and coupling function (\ref{cfunct}), gives

\begin{equation}\label{ESGB_null}
8\pi(E_{\alpha\beta})\!_{_{_{S\!G\!B}}}n^{\alpha}n^{\beta} = - \frac{2Q^{2}}{r^{4}}
\end{equation}
which is negative in all the WH domain as required. At  $r=r_{0}$, this is given by

\begin{equation}\label{ESGB_null_r0}
8\pi(E_{\alpha\beta})\!_{_{_{S\!G\!B}}} n^{\alpha}n^{\beta}\Big|_{r=r_{0}}  =  - \frac{2}{ Q^{2} } < 0.
\end{equation}
If we change the radial coordinate to $\rho = \sqrt{ r^{2} - Q^{2} }$, which implies $d\rho = (r/\sqrt{ r^{2} - Q^{2} })dr$, the line element (\ref{NewSolution_maxwell}) takes the simpler form,

\begin{equation}\label{genr_Ellis}
ds^{2} =  - dt^{2} +  d\rho^{2}  + (\rho^{2} + Q^{2} )(d\theta^{2}  + \sin^{2}\theta d\varphi^{2}).
\end{equation}
This is the same form of the metric originally obtained by Ellis, in  Ref. \cite{ellis73}, as an exact solution of Einstein gravity minimally coupled to a massless imaginary scalar field.
Therefore, in our case we have shown that the Ellis wormhole spacetime is also an exact solution of the EsGB-Maxwell theory,  but now with a real scalar field given by (\ref{phi_sol}) and a real coupling function given by (\ref{cfunct}).

\begin{figure}
\centering
\epsfig{file=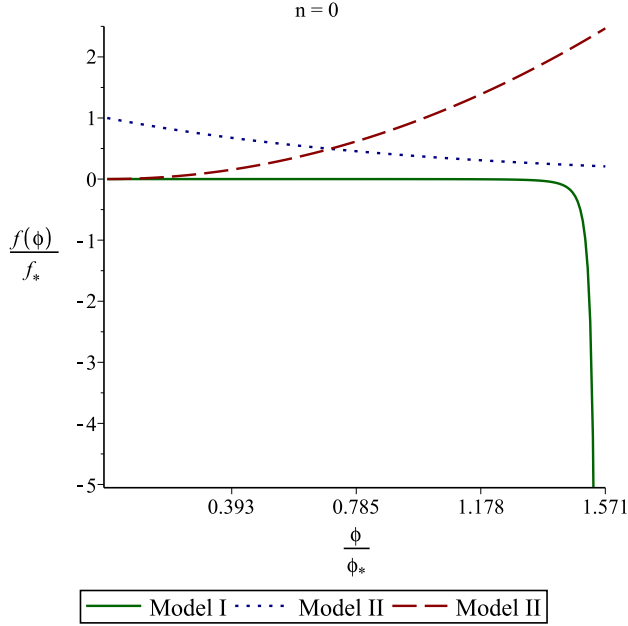, scale=0.44}
\caption{\label{fig1} The coupling functions $f(\phi)$ for different EsGB theories that admit traversable wormhole solutions are illustrated, for the branch $n=0$ defined in (\ref{rama}). The parameters $f_{\ast}$ and $\phi_{\ast}$,
are given respectively by $f_{\ast} = 10^{3} (Q^{2} + Q^{2}_{s})$ and $\phi_{\ast}=2Q_{s}/Q$. Model I, correspond to $f(\phi)$ given by (\ref{cfunct}); model II, $f(\phi) / f_{\ast} =  e^{-\phi/\phi_{\ast}}$;  model III, $f(\phi)/f_{\ast} = ( \phi / \phi_{\ast} )^2$.  Note that the range of the scalar field is $\phi_{0}\leq \phi < \phi_{\infty}$.
}
\end{figure}

\begin{figure}[h]
\hfill
\begin{minipage}[t]{.45\textwidth}
\begin{center}
\epsfig{file=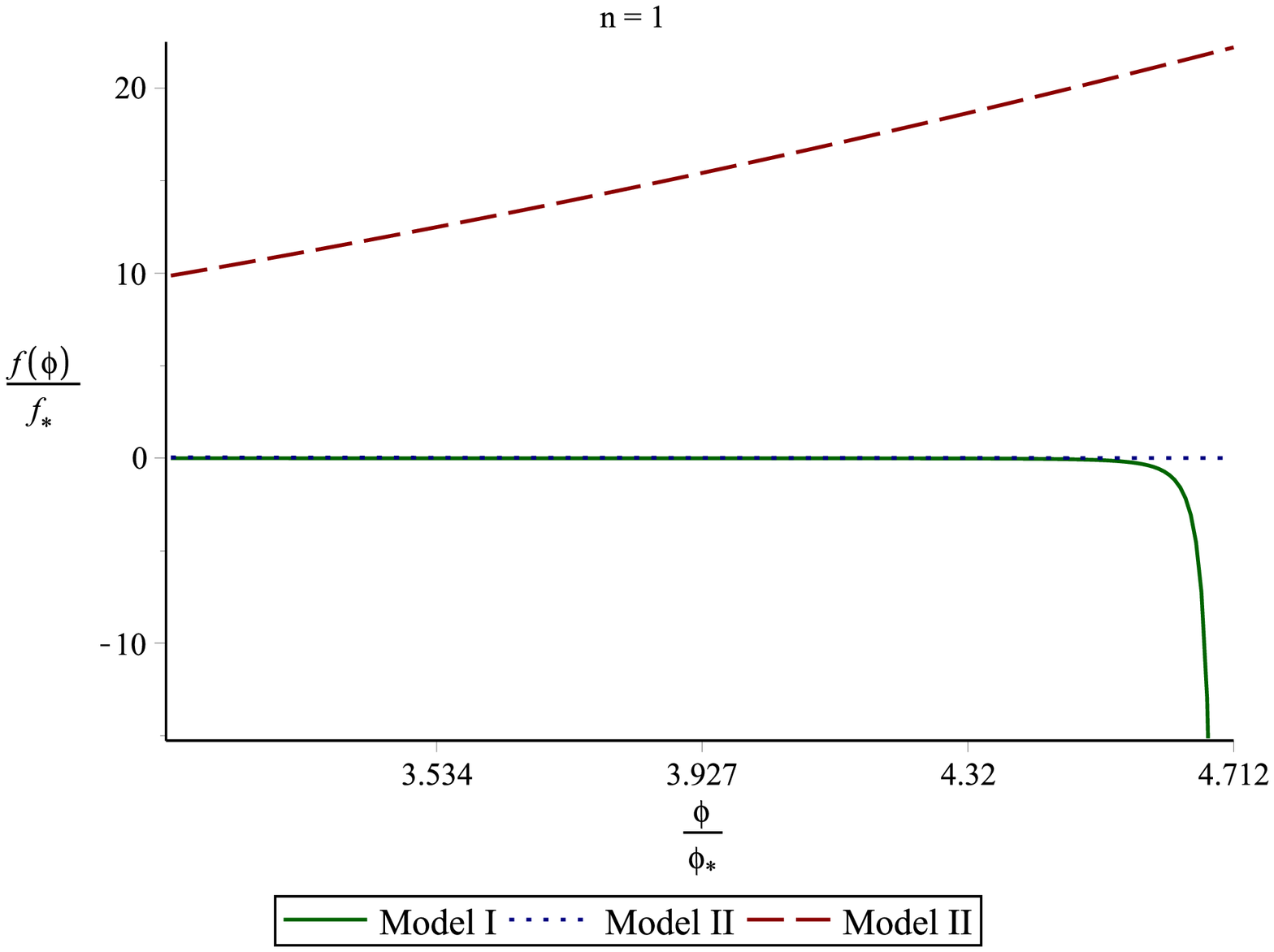, scale=0.44} 
\end{center}
\end{minipage}
\hfill
\begin{minipage}[t]{.45\textwidth}
\begin{center}
\epsfig{file=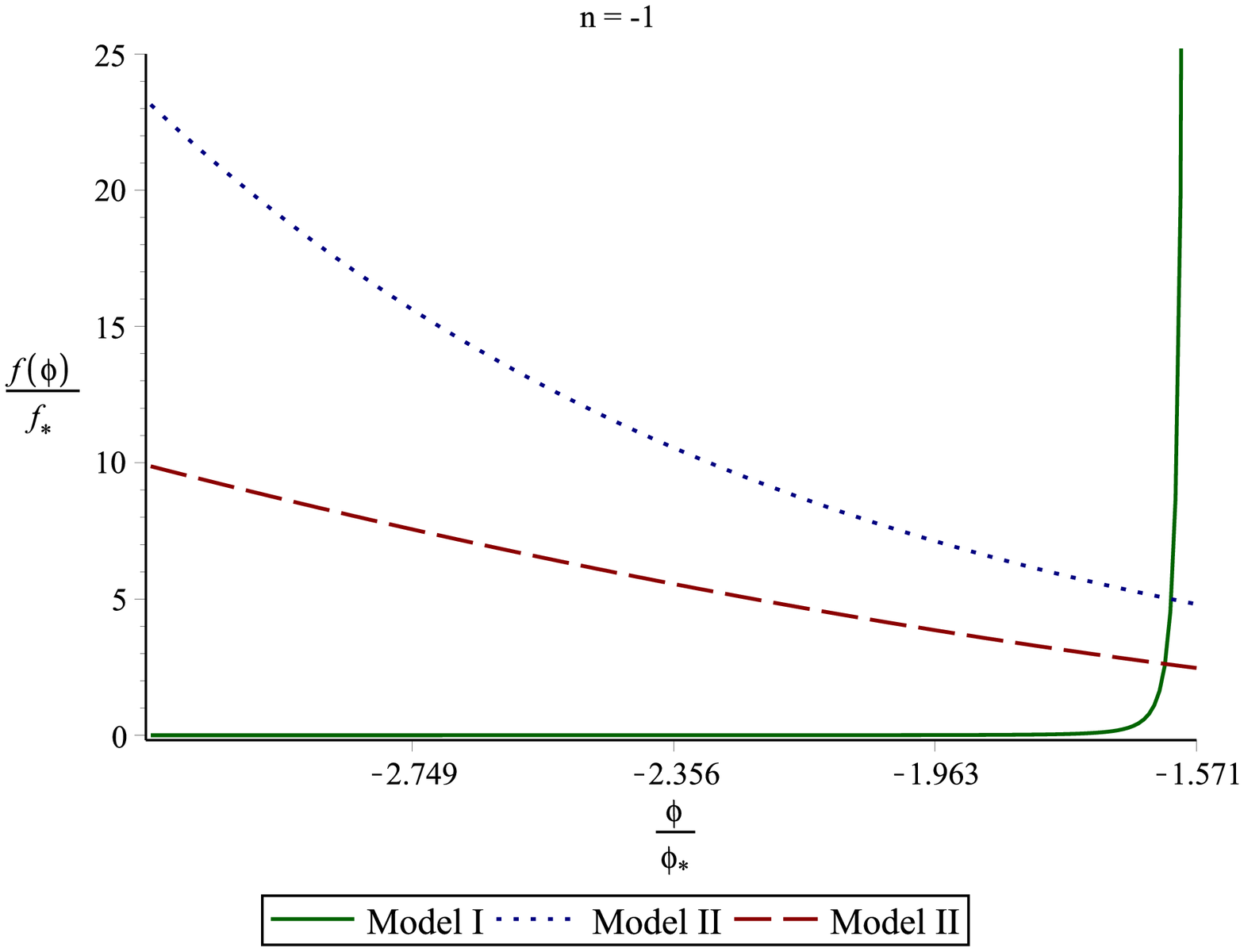, scale=0.43} 
\end{center}
\end{minipage}
\hfill
\caption{Similar as Fig. \ref{fig1}, with $n=1$ in the plot to the left; and $n=-1$ in the graphic to the right.}
\label{fig2}
\end{figure}

{\bf Other types of coupling functions.} Now, we focus on some coupling functions, $f(\phi)$, that like (\ref{cfunct}), also  admit traversable wormhole solutions in EsGB.
Our main objective here is to compare the behavior of these coupling functions, for the real values of scalar field (\ref{phi_sol}) that determine the wormhole solution (\ref{NewSolution_maxwell}). 
The range of the scalar field is limited by its value at the wormhole throat ($r=r_{0}$), $\phi_{0}$, and the value to which it tends in the asymptotically flat region ($r\sim\infty$), $\phi_{\infty}$.  
According to Eq. (\ref{phi_sol}), since the function $\tan(x)$ is periodic, then $\phi$ is not uniquely defined in terms of $r$; i.e., for each value of $r$ there are multiple values of $\phi(r)$, $\phi(r) + \frac{2Q_{s}}{Q} n\pi$, with $n\in \mathbb{Z}$. Thus, the values $\phi_{0}$ and $\phi_{\infty}$ are given respectively by 

\begin{equation}\label{rama}
 \phi_{0} = \phi(r_{0}) =  \frac{2 Q_{s}}{Q} n \pi, \quad \textup{ and } \quad
 \phi_{\infty} = \phi(r)\Big|_{r\sim\infty} =
 \frac{2Q_{s}}{Q} \left( \frac{\pi}{2} + n\pi \right), 
\end{equation}
where $n$ is the integer defining each particular branch.
Therefore, we will compare the behavior of the coupling functions in the domain $\phi\in[\phi_{0},\phi_{\infty})$. For simplicity, one can rename $\phi_{\ast} = \frac{2 Q_{s}}{Q}$. 
In Refs. \cite{antoniou19,kanti11} novel wormhole solutions were obtained, by using numerical methods,
in pure EsGB gravity with coupling functions having the form $f(\phi) = f_{\ast} e^{-\phi/\phi_{\ast}}$ and $ f(\phi) = f_{\ast} \left( \frac{ \phi }{ \phi_{\ast} } \right)^{2}$ respectively, where $f_{\ast}$ a positive real parameter.
However, despite the fact that in our treatment, in contrast to \cite{antoniou19,kanti11}, there is also an important contribution due to Maxwell electrodynamics, it is worth examine how much these $f(\phi)$-models resemble or differ from each other.
The results of the comparison of these $f(\phi)$-models are summarized in Fig. \ref{fig1} for $n=0$, Fig. \ref{fig2} for $n=1$, and $n=-1$, respectively. 

 \subsection{Ellis wormhole in Einstein-scalar-Gauss-Bonnet with trivial scalar field}\label{whEGBMax}
According to (\ref{phi_sol}) the trivial case of vanishing scalar field, $\phi(r)=0$, is obtained when setting $Q\!_{_{s}}=0$. For this  case the line element (\ref{NewSolution_maxwell}), and its traversable charged wormhole interpretation, are not modified.
However,  $Q\!_{_{s}} = 0$ also implies that $\phi' = 0$, $\phi'' = 0$, $\boldsymbol{f}(\phi(r)) \in \mathbb{R}$, $\dot{\boldsymbol{f}}(\phi) \rightarrow \infty$, and $\ddot{\boldsymbol{f}}(\phi) \rightarrow \infty$. Despite this, the quantities $\dot{\boldsymbol{f}}(\phi)$ and $\ddot{\boldsymbol{f}}(\phi)$, appear in the field equations (\ref{Eqt}), (\ref{Eqr}) and (\ref{Eqte}), as multiplicative factors to the derivatives of the scalar field, i.e., $\phi'\dot{\boldsymbol{f}}(\phi)$, $\phi''\dot{\boldsymbol{f}}(\phi)$ and $\phi'^{2}\ddot{\boldsymbol{f}}(\phi)$, and these are nonzero and regular in the wormhole domain and are given by

\begin{equation}
 \phi'\dot{\boldsymbol{f}}(\phi) = - \frac{ Q^{2}r^{3} \tan^{-1} \left(  \frac{\sqrt{ r^{2} - Q^{2}   }}{Q} \right)  }{ 2 Q^{3}   \sqrt{ r^{2} - Q^{2}   } }, \quad \phi''\dot{\boldsymbol{f}}(\phi) = - \frac{ Q^{2}\left( Q^{2} - 2r^{2} \right)r^{2} \tan^{-1} \left( \frac{\sqrt{ r^{2} - Q^{2}   }}{Q} \right)  }{ 2 Q^{3}  \left( r^{2} - Q^{2} \right)^{\frac{3}{2}}  }
\end{equation}
and

\begin{equation}
 \phi'^{2}\ddot{\boldsymbol{f}}(\phi) = \frac{ Q^{2}r^{2} \left( 4\sqrt{ r^{2} - Q^{2}   }\tan^{-1} \left( \frac{\sqrt{ r^{2} - Q^{2}   }}{Q} \right) + Q \right) }{ 2 Q^{3}  \left( Q^{2} - r^{2} \right)  }.
\end{equation}
 Moreover, for this case the scalar field $\phi$ and the coupling function $\boldsymbol{f}$ are decoupled, in such way that $\boldsymbol{f}$ as function of radial coordinate takes the form,

\begin{equation}
 \boldsymbol{f}(r) = \frac{r^{2}}{12} - \frac{Q^{2}}{12} \ln\left(\frac{ Q^{4} }{ r^{4} }\right) - \frac{1}{3}\left(Q  + \frac{r^{2}}{2Q}\right)\sqrt{ r^{2} - Q^{2}   }\tan^{-1} \left( \frac{\sqrt{ r^{2} - Q^{2}   }}{Q} \right).
\end{equation}

Summarizing, we can say that when $Q\!_{_{s}}=0$, the scalar field becomes trivial, $\phi(r)=0$. However, the line element (\ref{NewSolution_maxwell}), together with the  electromagnetic invariant (\ref{Fsol}), electromagnetic  Lagrangian $\tilde{\kappa} F$ with parameter $\tilde{\kappa}$ given by (\ref{kappa}), scalar field and coupling function given respectively by (\ref{phi_sol})  and (\ref{cfunct}), with $Q\!_{_{s}} = 0$, is still a  solution of (EsGB-Maxwell).  In this case the parameter $Q$ in the line element (\ref{NewSolution_maxwell}) represents the electric charge and the traversable Ellis wormhole is only supported by the source-free electric field and the nonzero coupling of the GB term.

\subsection{Ellis wormhole without electric charge: Original Ellis work}
Equation (\ref{Fsol}) indicates that if the parameter $Q\!_{_{s}}$ is imaginary and such that $Q\!_{_{s}} = i Q$, then the electromagnetic invariant vanishes, $F(r)=0$. Thus, for this value of the parameter $Q\!_{_{s}}$, there is no electromagnetic contribution to the traversable wormhole spacetime (\ref{NewSolution_maxwell}). Thus, in this case, the parameter $Q$ in the wormhole metric (\ref{NewSolution_maxwell}), can be associated solely with the scalar field (\ref{phi_sol}); i.e., $Q$ is interpreted as the charge of the scalar field. Now, for $Q\!_{_{s}} = i Q$ the scalar field is imaginary and is given by

\begin{equation}\label{phi_sol_phant}
\phi(r) =   2i\tan^{-1}\left( \frac{ \sqrt{r^{2} - Q^{2}} }{Q} \right).
\end{equation}
Furthermore, according to the Eq. (\ref{cfunct}),  when setting $Q\!_{_{s}}=i Q$ the coupling function $\boldsymbol{f}(\phi)$ and all its derivatives with respect to the scalar field become zero. \\[2mm]
So defining a new scalar field by $\psi = i\phi$ (phantom field), the wormhole metric (\ref{NewSolution_maxwell}) becomes a solution to the theory with gravitational action,

\begin{equation}\label{actionL2}
S[g_{ab},\psi] = \int d^{4}x \sqrt{-g} \left\{ \frac{1}{16\pi}\left(R + \frac{1}{2}\partial_{\mu}\psi\partial^{\mu}\psi \right)  \right\},
\end{equation}
with $\psi$  given by

\begin{equation}
\psi =  2\tan^{-1}\left( \sqrt{ \frac{r^{2}  - q^{2}}{q^{2}} } \right)  \in \mathbb{R}.
\end{equation}
This is the action that was used by Ellis in Ref. \cite{ellis73} to get the wormhole solution (\ref{NewSolution_maxwell}).

\section{Conclusion}

In this paper we have shown that the traversable Ellis wormhole is a solution of EsGB-Maxwell gravity with a material source consisting of source-free electric field and a real scalar field. In general relativity this solution requires a source with negative energy density as shown in the original derivation by Ellis \cite{ellis73} and in the recent interpretation by Bronnikov {\it et al.} \cite{bronnikov13}. Hence unlike these previous results, in our case the violation of the null-energy condition in (\ref{ESGB_null}) required for traversability of the wormhole, arises from the theory itself, and so it is of purely gravitational nature.\\[2mm]
Like many other exact wormhole solutions found in the literature, the Ellis wormhole has been shown to be unstable \cite{EllisStability}. However as pointed out by Bronnikov {\it et al.} in Ref. \cite{bronnikov13} the stability of a solution, in general depends also on its source, i.e. on the associated energy momentum tensor. In fact they showed that their interpretation of the Ellis wormhole as a solution in general relativity with a material source made up of a combination of a source-free electric field and phantom dust, makes the solution linearly stable under both spherical and polar perturbations. So inspired by this study, in a future publication we intend to examine the stability of the Ellis wormhole as a solution to the EsGB-Maxwell gravity.\\

\textbf{Acknowledgments}: P. C. acknowledges partial financial support from CONACYT-Mexico through the Project No. 284489; as well as thanks the Physics Department of Cinvestav for hospitality.

\end{document}